\documentclass[english,aps,prper,reprint,showpacs,titlepage,longbibliography]{revtex4-2}   

\usepackage[T1]{fontenc}	
\usepackage{geometry}
\usepackage{graphicx}
\usepackage{times}
\usepackage{hyperref}  
\hypersetup{colorlinks=true,urlcolor=blue,citecolor=blue,linkcolor=blue}
\usepackage{array}
\usepackage{enumerate}
\usepackage{enumitem}  
\usepackage{amsmath}
\usepackage{amssymb}
\usepackage{tikz}
\usepackage{graphicx}
\usepackage{multirow}
\usepackage{tcolorbox}
\usepackage{ragged2e}
\usepackage{tcolorbox}
\usepackage{float}
\usepackage[utf8]{inputenc}

\usepackage[normalem]{ulem}

\begin{document}
\begin{titlepage}

\title{Dual-Role Dynamics in Prompting: Elementary Pre-service Teachers’ AI Prompting Strategies for Representational Choices}

 \author{Razan Hamed}
 \affiliation{Department of Physics and Astronomy, Purdue University, 525 Northwestern Ave, West Lafayette, IN, 47907, U.S.A.} 
 
 \author{Amogh Sirnoorkar}
 \affiliation{Department of Physics and Astronomy, Purdue University, West Lafayette, IN, 47907, U.S.A.\\
 Department of Curriculum and Instruction, Purdue University, West Lafayette, IN, 47907 U.S.A.} 

 \author{N. Sanjay Rebello}
 \affiliation{Department of Physics and Astronomy, Purdue University, West Lafayette, IN, 47907, U.S.A.\\
 Department of Curriculum and Instruction, Purdue University, West Lafayette, IN, 47907 U.S.A.} 
 

\begin{abstract}

Pre-service teachers play a unique dual role as they straddle between the roles of students and future teachers. This dual role requires them to adopt both the learner's and the instructor's perspectives while engaging with pedagogical and content knowledge. The current study investigates how pre-service elementary teachers taking a physical science course prompt AI to generate representations that effectively communicate conceptual ideas to two distinct audiences. The context involves participants interacting with AI to generate appropriate representations that explain the concepts of wave velocity to their elementary students (while casting themselves as teachers) and the Ideal Gas Law to their English teachers (while casting themselves as students). Emergent coding of the AI prompts highlight that, when acting as teachers, participants were more explicit in specifying the target audience, predetermining the type of representation, and producing a broader variety of representations compared to when they acted as students. Implications of the observed `exploratory' and `prescriptive' prompting trends across the two roles on pre-service teachers' education and their professional development are discussed. 

\clearpage
\end{abstract}

\maketitle
\end{titlepage}


\section{Introduction}
\label{sec:intro}

Pre-service teacher training has been one of the key focal areas in science education research~\cite{azra2025going} as it is crucial for shaping the next-generation science learning. However, the training involves inherent complexities, as these pre-service teachers actively engage not only with content knowledge but also with pedagogical practices~\cite{gess1999pedagogical} and pedagogical content knowledge \cite{shulman1986pedagogical}. Since the candidates often simultaneously teach while taking science courses, they continually shift between the roles of a teacher (when teaching in classrooms) and that of a student (when enrolled in science courses). Given the influence of pedagogical environments on learning approaches~\cite{stroupe2019introduction}, it becomes vital to examine pre-service teachers’ learning experiences in science, vis-\`{a}-vis their dual roles as teachers and as students.

One of the ways in which education researchers have explored students’ learning is by looking at the questions that students generate ~\cite{chin2002student}. Student-generated questions assume significance as they are an essential component of discursive activity and dialectical thinking, both central to knowledge construction~\cite{chin2008students}. Students' vexing questions can lead them to notice gaps in their understanding thereby leading them to engage in valued cognitive processes such as sensemaking and  modeling~\cite{odden2019vexing,sirnoorkar2023sensemaking}. The centrality of questioning in science education is evident from ``Asking Questions'' being identified as one of the key scientific practices to be promoted in classrooms~\cite{national2012framework}. 

Recent advancements in Generative-Artificial Intelligence (henceforth referred to as ``AI'') have opened up new avenues to facilitate science learning, particularly for pre-service teachers. The increasing focus on AI comes against the backdrop of its ability to generate multi-modal, coherent, and context-specific responses based on the user input~\cite{sirnoorkar2024student,bralin2024}. Studies have observed AI's potential to facilitate prospective teachers’ ability to design tasks, develop lesson plans, practice Socratic questioning, etc. (Refer Section~\ref{sec:background} for additional details). However, research on pre-service teachers’ questioning approaches in a physics-based context, particularly while interacting with AI remains underexplored. Furthermore, given the complexity of pre-service teachers’ dual roles in their professional journey (as instructors in elementary schools and as students taking courses), it becomes imperative to explore the nature of their questions and prompts to AI while taking on each of the two roles.

We address the above gaps in the literature by investigating pre-service teachers’ questions in the form of prompts to AI while determining the appropriate representations that explain the concepts of (i)  wave velocity to their elementary students (as a teacher)  and (ii) the Ideal Gas Law to their English teacher (as a student). In doing so, we address the following research question:{\em How do the dual roles of pre-service elementary teachers' -- their role as teachers and their role as students -- influence their AI prompts in the context of choosing effective representations in physics?}
In the next section, we provide background literature on pre-service teachers in the context of using AI. In Section~\ref{sec:methods}, we detail the study’s methodology. In Section~\ref{sec:results}, we present the results before discussing their implications in Section~\ref{sec:discussion}.

\section{Background}
\label{sec:background}

 The recent advancements in AI have resulted in researchers exploring its potential use in training future teachers. Studies exploring pre-service teachers’ perspectives on using AI have highlighted that they recognize a strong potential for incorporating AI into science teaching and learning. However, many pre-service teachers tend to lack a clear understanding of AI tools’ functionalities and capabilities~\cite {lee2024using}. Thus, to build pre-service teachers’ AI literacy and foster a deeper understanding of AI's applications in education, it is crucial to integrate a range of AI tools into their professional development programs\cite{ayanwale2024exploring}. Such integration will equip them with the technical knowledge and pedagogical skills needed to tackle potential challenges in their classrooms~\cite{abdulayeva2025fostering}. 

A study targeting how pre-service elementary teachers frame their questions has shown that pre-service teachers tend to ask their students short fast-paced questions that are based on task completion rather than exploration, lacking any follow up questions \cite{moyer2002learning}. AI tools like ChatGPT, given their impressive abilities to contextualize and adapt to user prompts, have promised potential in helping pre-service teachers facilitate more adaptive and diverse questions \cite{erdem2025artificial}. Beyond enhancing communication competencies, AI also has the potential to reshape instructional methods and improve pre-service teachers' pedagogical adaptability by assisting them in generating innovative ways to deliver learning materials \cite{siddiquiai}. Moreover, AI tools have been shown to enhance pre-service teachers’ confidence, engagement, and readiness for science teaching \cite{ayanwale2024exploring}.

As learners themselves, pre-service teachers benefit greatly from AI's personalized tutoring that supports them based on their background knowledge and personal preferences \cite{hwang2014definition}. Similarly, within the realm of physics education, AI has shown a significant potential to enhance physics learning across all educational levels \cite {shafiq2025artificial}. Due to the complex nature of physics concepts and the need for practical applications, AI-powered tools enable students to explore such concepts dynamically and intuitively fostering a deeper level of comprehension \cite{mahligawati2023artificial}. One of the issues students tend to struggle with is the use of mathematical functions in physics, for which AI tools derive symbolic expressions to help students understand the relationships between variables in complex physics problems \cite{verawati2024role}. 

Like any new technology, AI has its own limitations; while AI can enhance the personalization of the learning experience, over-reliance on AI risks reducing creativity, critical thinking, and originality among pre-service teachers \cite{bae2024pre}. Therefore, it is important to recognize the consequences of using AI while maximizing the benefits and the advantages AI offers for improving teaching and learning. Incorporating AI literacy programs into teacher education curricula has been shown to foster effective use of AI while being critical of its outcomes and feedback \cite{abdulayeva2025fostering}. Thus, comprehensive AI literacy training ensures that AI is used as a supportive tool in education rather than a substitutive one. 

If used properly, AI can offer significant support for pre-service teachers in their teacher education in general and in their physics education in specific. Given their dual role, pre-service teachers experience the effects of AI both as students considering their own learning, and as teachers assisting in-service teachers to complete their professional training. A previous study exploring the effect of pre-service teachers' dual role on using AI indicated that they intend to utilize AI tools for writing tasks and developing study plans as students, while using AI tools in lesson planning and brainstorming as teachers \cite{bae2024pre}. However, there is little to no research on the dual role of pre-service teachers in physics education or their prompting approaches in seeking appropriate representations. Therefore, this study bridges the gap by exploring how pre-service teachers interact with AI to generate physics representations as they step into the role of a teacher compared to that of a student.

\begin{figure}[tb]
    \centering
    \begin{tcolorbox}
    \noindent
     \justify{You are a science teacher tasked with teaching elementary students (K-6) about waves. Your goal is to effectively communicate complex physics ideas to your students using appropriate representations, particularly about wave velocity, in a manner that your students would understand. (If you need help with your understanding about waves or wave velocity, feel free to use any resources you would like (e.g., lecture slides, internet, etc.), but DO NOT USE AI).
     
     Choose your students grade level:  
     
     1. How would you effectively present the concept of wave velocity to your students? Use two of the following representations in your explanations: pictures, cartoons, equations, analogies, graphs, animations, tables containing numbers (data), stories, scripts, plays, games, or drawings.
    
    2. Repeat everything you have done in part 1, but now using AI (preferably Co-pilot but you are welcome to use any other AI tool) [Link to Co-Pilot]} 
    
\end{tcolorbox}
\caption{Teacher Role. Problem statement on wave velocity where participants used AI as teachers to come up an appropriate representation to effectively communicate the concept to their elementary students. }
\label{fig:problem-statement-teachers}
\end{figure}

\section{Methods}
\label{sec:methods}


Participants in our study were enrolled in a physical-science course that is required for elementary education majors at a large midwestern land-grant R1 university. The course consists of two laboratory sessions and one lecture every week. It is organized into  four units: Mechanics,  Circuits, Waves, and Thermodynamics. This study was conducted in Spring 2025 and involved Waves and Thermodynamics units. The enrollment was 34 elementary education majors. All participants were simultaneously engaged in field experiences to shadow in-service teachers in local elementary schools, although these field experiences were not part of this course. The data for this study were collected over two laboratory sessions in which the participants assumed either the role of a teacher (in Session 1) or that of a student (in Session 2). Session 1 took place at the end of the 'Waves' unit with extensive focus on waves and their properties. The participants assumed the role of teachers in this lab, as they had exposure to the relevant content particularly `Wave velocity'. Session 2 on the other hand, took place before the beginning of the `Thermodynamics' unit and because the participants had not received formal instruction on the topic at this point, they were well-positioned to reason as students. The topic of interest in this lab was `Ideal Gas Law'.

\begin{figure}[tb]
    \centering
    \begin{tcolorbox}
    \noindent
     \justify{After learning about the Ideal Gas Law, you come across your English teacher in the hallway who asks you what you have been up to. You tell them you have been exploring the Ideal Gas Law, to which they asked what the law was.  

    1. How would you explain the Ideal Gas Law to your English teacher? Use two of the following representations in your explanations: pictures, cartoons, equations, analogies, graphs, animations, tables containing numbers (data), stories, scripts, plays, games, or drawings.

    2. Now you are interested in how AI would come up with the best representation to explain the Ideal Gas Law to your English teacher (It is preferred to use Co-pilot here, but you are welcome to use any other AI tool). [Link to Co-Pilot] }

\end{tcolorbox}
\caption{Student Role. Problem statement on Ideas Gas Law where participants used AI as students to come up an appropriate representation to effectively communicate the concept to their English teacher.}
\label{fig:problem-statement-students}
\end{figure}

In session 1 (henceforth referred to as `Teacher Role'), participants were asked to come up with a representation initially without using AI, and later with the help of AI specifically using \emph{Microsoft Co-Pilot Pro} to effectively explain the concept of wave velocity to their elementary students. In session 2 (henceforth referred to as `Student Role'), the participants were {\em expected} to act as students given their limited exposure to the topic of interest, the Ideal Gas Law. Given the lack of content familiarity, session 2 differed slightly from session 1. At the beginning of session 2, participants were given three real-world scenarios that involved the interplay between pressure, volume, and temperature to set the stage for the Ideal Gas Law. Participants were asked to use all available resources to identify the interplay between the physical quantities and arrive at the Ideal Gas Law. Similar to session 1, participants were then instructed to come up with a representation to explain the Ideal Gas Law to their English teacher initially without, and later with the help of AI (Microsoft Co-Pilot Pro). In both roles, to better facilitate their engagement with the activity, participants were given a set of representations to choose from and were also encouraged to come up with their own representations. They were also explicitly told not to copy and paste the questions while interacting with AI, but rather converse in a manner that suits them best using their own words. Figures~\ref{fig:problem-statement-teachers}~and~\ref{fig:problem-statement-students} represent the wording of the activities across the two labs.    



\renewcommand{\arraystretch}{1.1}
\begin{table*}[tb]
\begin{ruledtabular}
\caption{\label{tab:prompt-examples} Exemplar AI prompts of physics pre-service teachers while in the roles of a teacher and a student.}

\begin{tabular}{p{0.14\linewidth} p{0.4\linewidth} p{0.4\linewidth}}

{\bf Feature exemplars}  & {\bf Teacher role } & {\bf Student role} \\ 
\hline 

Prompt specifying target audience  & ``{\em Out of these methods: pictures, cartoons, equations, analogies, graphs, animations, tables containing numbers (data), stories, scripts, plays, games, or drawings. Which method would you use to explain wave velocity to 5th graders?}'' & ``{\em Can you give me an analogy for explaining the gas law to my English teacher?}''  \\

Prompt without target audience  & ``{\em Can you design a graph to represent wave velocity?''} & ``{\em Create a picture that represents gas molecules being affected by temperature using volume, temperature, and pressure.} \\

Prescriptive prompt & ``{\em What is a photo that shows wavelength but in waves in the ocean for elementary school?}'' & ``{\em What’s a simple analogy I can use to explain the Ideal Gas Law to a non-scientist?}'' \\

Exploratory prompt & ``{\em What way should I represent the concept of wave velocity to 4th graders?}'' & ``{\em What is the best representation to explain the Ideal Gas Law to your English teacher?} \\

Representations & Picture, graph, cartoon, analogy, and activity. & Picture, analogy, equation, and story.   \\

\end{tabular}
\end{ruledtabular}
\end{table*}

The participants' prompts and AI responses were then collected to explore their interaction with AI across the two roles. Of the total 34 participants enrolled in the course, 28 completed session 1 and 24 completed session 2. To ensure consistency across the data, we selected only those participants' responses: (i) who attended both sessions and (ii) responded to the specific questions on AI that are relevant to this study. These criteria narrowed down the sample size to 20 participants. Each of the participants' prompts along with the corresponding AI responses across the two sessions were then organized in a spreadsheet. Qualitative approaches were employed to analyze the data, particularly the participants' prompts. 



\section{Results}
\label{sec:results}

The objective of this study is to empirically explore the features of questions posed by elementary education majors to Generative-AI inquiring about the appropriate representations to communicate wave velocity and Ideal Gas Law concepts by stepping into the role of both teachers and students. Through emergent coding of participants' prompts, we identified three features that seemed to vary across the two roles: (i) explicit specification of the target audience, (i) predetermination of representation, and (iii) diversity of representations. Table~\ref{tab:prompt-examples} provides exemplar prompts to better contextualize the results discussed below.




The first emergent feature across the two roles corresponds to explicit specification of the target audience: elementary students in the teacher role and English teachers in the student role. While 17 participants specified the target audience in the teacher role, only 11 participants did so in student role. Participants’ exemplar prompts with explicit mention of (as well as lack of) target audiences from each role are given in Table~\ref{tab:prompt-examples}. While nine participants specified the target audience in both roles, eight did so in the teacher role but not in the student role. However, in only two cases the opposite was observed, i.e., representation specified in student role but not specified in the teacher role. Only one participant did not specify the audience in both roles. These results highlight that participants were relatively more specific in mentioning the target audience explicitly in their prompts when acting as teachers as compared to students.

\renewcommand{\arraystretch}{1.2}







The second emergent feature corresponds to the extent of certainty about the representation type specified in the prompts. 
While few participants predetermined the appropriate representation and instructed AI to generate their chosen one, others sought and allowed AI to determine the appropriate representation for them. We refer to the former  prompt categories as “prescriptive” and the latter as “exploratory”. The data reflected 17 out of 20 participants’ prompts to be prescriptive in the teacher role as compared to 11 in the student role. 
While 11 participants were prescriptive in both roles, six were prescriptive in the teacher role but not in the student role.  There were however no prompts that were exploratory in the teacher role but prescriptive in the student role. Only three participants prompted exploratory questions in both roles. In summary, we observe that in the role of teachers, participants’ AI prompts are more likely to be prescriptive with pre-determined representations. However, in the role of students, their prompts are likely to be more exploratory by letting AI to decide the appropriate representations.

The last emergent feature corresponds to the diversity of representations generated across the two roles. Participants across the two roles generated diverse representations, few of which were characteristic to each role. In the teacher role, participants generated pictures (9), cartoons (2), analogies (3), activity (1), and graphs (2). In the student role, participants generated pictures (2), analogies (4), story (1), and equations (5). Only three of the 20 participants generated the same representation type across the two roles. While cartoons, activities, and graphs were unique representations generated in the teacher role, stories and equations were uniquely generated in the student role.  Thus, we observed elementary majors tend to generate relatively more diverse representations when positioned as teachers as compared to students.

\section{Discussion and Conclusion}
\label{sec:discussion}

In the previous section, we discussed the emergent features of elementary pre-service teachers' prompting to AI while being in the role of a teacher and that of a student. The context involved seeking appropriate representation from AI to better communicate the concepts of wave velocity to their elementary students (in the teacher role) and the Ideal Gas Law to their English teachers (in the student role). The results are summarized below:

\begin{enumerate}
    \item Participants were relatively more explicit in mentioning the target audience in their prompts when acting as teachers as compared to students. 

    \item The prompts were more likely to be prescriptive with pre-determined representations when acting as teachers. When acting as students, the prompts were more likely to be exploratory (seeking representations).

    \item The participants tended to come up with relatively more diverse representations through AI while being in the role of a teacher than that of  a student.      
\end{enumerate}  

Our study with the above results makes several key contributions to the current understanding about pre-service teachers’ prompting of AI. Firstly, though several studies have explored pre-service teachers’ interactions with AI, very few have explicitly focused on comparing the prompting patterns while ‘wearing the cap’ of a teacher versus that of a student. Our study contributes to this literature by focusing on the AI prompts by being in one of the two roles. Secondly, though contemporary studies have explored pre-service teachers’ use of AI in contexts such as lesson planning~\cite{moorhouse2025pre,kerr2025prompts} and task design~\cite{kuchemann2023physics}, this study focuses on the context of generating representations to better communicate conceptual ideas. Focusing on representations assumes significance as one of the key strengths of AI lie in its multimodality, i.e., its ability to engage with multiple input and output representational formats. Furthermore, `representational competence’~\cite{daniel2018towards} is also a key component of science education.

Findings reported in this study have several implications for educators. Our observations that pre-service teachers are relatively more explicit and prescriptive in their prompts to AI, indicate at the outset, a relatively higher degree of agency when engaging with content while adopting the role of a teacher as compared to that of a student. Teacher preparation courses and programs can thus tap into this potential to better facilitate pre-service teachers’ content learning. Pre-service teachers should be actively encouraged to approach their learning in science courses from the perspective of a teacher who is capable of comprehending the learning material beyond what they would need to deliver to their students.  Furthermore, our observations that these pre-service teachers' tend to generate more diverse representations when acting as teachers can further inform the design of science content courses for pre-service teachers. These courses can actively encourage pre-service teachers to create multiple representations of science concepts to become more confident learners and successful educators. This goal is made even more achievable with the assistance of AI that has the added capability to engage with multi-modal data.

The findings also accompany several limitations. Firstly, the small sample size of 20 participants. Though consistent with contemporary studies with a similar research focus~\cite{moorhouse2025pre}, observations made in this study are limited by the small sample size. Secondly, the nature of activity that the participants engaged in, i.e., generating appropriate representations of a concept was very linear and specific focusing on only two types of audiences. More iterative and longitudinal activities focused on the dual role across diverse content areas and a wider range of audience can potentially shed additional insights onto our observations. Lastly, to minimize any bias stemming from the order of using AI in the two lab sessions, a between-subject study could yield noteworthy outcomes. 

Future work would focus on addressing these limitations along with explicitly exploring the interactions through the theoretical lenses of `agency' and `representational competence'. Agency corresponds to the ability to exercise influence over one’s own circumstances in a causal manner where the person is an active contributor to their environment rather than simply being a product of it \cite{bandura2006toward}. Representational competence, on the other hand, represents  skills and practices that allow a person to reflectively use multiple forms of representations to communicate domain-specific concepts effectively \cite{kozma2005students}. Rather than being emergent themes arising from the data, both concepts will serve as foundational anchors in future studies to guide the analysis of pre-service teachers' prompting behavior and question framing while interacting with AI. Another avenue for future work includes investigating pre-service teachers approach and willingness to ask AI follow up questions when generating representations in physics. Such research would target the implications of asking short convergent questions as opposed to exploratory conversation-based questions while interacting with AI.

\section{Acknowledgments}
ChatGPT-4.0 was employed only to address grammatical errors, but not for  ideation or analysis. This research is partly supported by Purdue University's Innovation Hub.
\clearpage
\bibliography{references}

\end{document}